\documentclass[%
 reprint,
 amsmath,amssymb,amsthm
 aps,
]{revtex4-1}

\usepackage{array}
\usepackage{stmaryrd}
\usepackage{graphicx}
\usepackage{dcolumn}
\usepackage{bm}

\newtheorem{definition}{Definition  }[section]
\newtheorem{theorem}{Theorem}

\newtheorem{ex}{Example     }[section]
\newtheorem{rem}{Remark    }[section]

\def\CC{\mathbb{C}}
\def\PP{\mathbb{P}}
\newcommand{\bra}[1]{{\left\langle{#1}\right\vert}}
\newcommand{\ket}[1]{{\left\vert{#1}\right\rangle}}

\begin{document}


\title{Three-qutrit entanglement and simple singularities}

\author{Fr\'ed\'eric Holweck}
\email{frederic.holweck@utbm.fr}
\author{Hamza Jaffali}%
 \email{hamza.jaffali@utbm.fr}
\affiliation{IRTES-UTBM, Universit\'e de Bourgogne Franche-Comt\'e,\\ 90010 Belfort Cedex, France
}%

%
%

\date{\today}

\begin{abstract}
In this paper, we use singularity theory to study the entanglement nature of pure three-qutrit systems. We first consider the algebraic variety $X$ of separable three-qutrit states within the projective 
	Hilbert space $\PP(\mathcal{H}) = \PP^{26}$. Given a quantum pure state $\ket{\varphi}\in \PP(\mathcal{H})$ we define the $X_\varphi$-hypersuface by cutting $X$ with a hyperplane $H_\varphi$  defined by the linear form $\bra{\varphi}$ 
	(the $X_\varphi$-hypersurface of $X$ is $X\cap H_\varphi \subset X$).
	We prove that when $\ket{\varphi}$ ranges over the SLOCC entanglement classes, the ``worst'' possible singular $X_\varphi$-hypersuface with isolated singularities, has a unique
	singular point of type $D_4$.
\begin{description}
\item[PACS numbers] 03.65.Ud,03.67.Hk,03.65.Aa,03.65.Db
\item[Keywords] Entanglement, three-qutrit systems, singularity theory.
\end{description}
\end{abstract}

\maketitle


\section{Introduction}




	Let $\mathcal{H} = \CC^3 \otimes \CC^3 \otimes \CC^3$ be the Hilbert space of pure three-qutrit states. Up to multiplication by a scalar, every 
	three-qutrit state $\ket{\psi} \in \mathcal{H}$ can be considered as a point of the projective space $\PP^{26} = \PP(\mathcal{H})$ ($\PP^N$ will denote a projective space of dimension $N$). Separable states of $\mathcal{H}$ are 
	states that can be factorized as $\ket{\phi} = v_1 \otimes v_2 \otimes v_3$, with $v_i \in \CC^3$. The set of separable states is indeed the set of tensors of rank 1. We denote by $\{\ket{0},\ket{1},\ket{2}\}$ 
	the single-qutrit computational basis and we  denote by $\ket{ijk} = \ket{i} \otimes \ket{j} \otimes \ket{k}$  the three-qutrit basis. Thereby, we  write a general three-qutrit state $\ket{\psi} \in \mathcal{H}$ as: 

\begin{equation} \ket{\psi} = \sum_{i,j,k\in \llbracket 0,2 \rrbracket} A_{ijk}\ket{ijk} \text{, where } A_{ijk} \in \CC \text{.}\end{equation}

	Let $G$ be the group of Stochastic Local Operation and Classical Communication (SLOCC) of three-qutrits (acting on $\PP(\mathcal{H})$), we have 
	$G = SL_3(\CC) \times SL_3(\CC) \times SL_3(\CC)$. The group $G$ acts transitively on the set of separable states.
	The corresponding orbit, which is the equivalence class of $\ket{000}$ for the equivalence relation defined by

\begin{equation}\forall ~ \ket{\psi_1},\ket{\psi_2} \in \PP(\mathcal{H}), ~ \ket{\psi_1} \sim \ket{\psi_2} \Leftrightarrow \exists g \in G ~~~~ \ket{\psi_1} = g.\ket{\psi_2}, \end{equation}

is also called the highest weight orbit. Its projectivization is the unique smooth orbit $X$ for the action of $G$ on $\PP(\mathcal{H})$, more precisely: 

\begin{equation}\label{seg} X = \PP(G.\ket{000}) = \{\text{The set of separable states}\} \subset \PP^{26}\text{.}\end{equation}

We can parametrize this unique smooth orbit $X$  using the Segre embedding \cite{Ha} of three projective planes: 

\begin{equation} \mathcal{S}: \PP^2 \times \PP^2 \times \PP^2  \to  \PP^{26}\end{equation} with \begin{equation}\label{segreeq}\begin{split} \mathcal{S}([x_0:x_1:x_2],[y_0:y_1:y_2],[z_0:z_1:z_2])\\= 
[x_0y_0w_0: \dots: W_J: \dots: x_2y_2z_2]\end{split}\end{equation}

where $W_J = x_i y_j z_k$ for $J = \{i,j,k\} \in \llbracket 0,2 \rrbracket ^3$. 

The monomials are ordered according to the value in base 3 of their indexes, such that $W_{J_1} \prec W_{J_2}$ iff $9i_1+3j_1+k_1 \leq 9i_2+3j_2+k_2$.

Let us introduce the notion of hyperplane to implement our construction. We denote by $\bra{\varphi}$, a linear form on $\mathcal{H}$ defined by the state $\ket{\varphi}$ via the inner product on the Hilbert space.
The hyperplane $H_\varphi \subset \PP(\mathcal{H})=\PP^{26}$ is the zero locus  of $\bra{\varphi}$, i.e. is the set of  states $\ket{\psi} \in \PP(\mathcal{H})$ such as 
$\langle\varphi|\psi\rangle = 0$. Then we construct the hyperplane section $H_\varphi \cap X$, that is the restriction of $\bra{\varphi}$ to $X$. 
In other words this hyperplane section is the set of separable states on which $\bra{\varphi}$ vanishes. Let $\bra{\varphi} = \sum_{i,j,k\in \llbracket 0,2 \rrbracket} h_{ijk} \bra{ijk}$ 
be the linear form defining the hyperplane $H_\varphi$. Then the hyperplane section $H_\varphi \cap X$ will be, according to the Segre map (see Eq. (\ref{segreeq})), the $X_\varphi$-hypersurface of $X$ given by

\begin{equation} \langle\varphi|\mathcal{S}(\PP^2 \times \PP^2 \times \PP^2)\rangle = \sum_{i,j,k\in \llbracket 0,2 \rrbracket} h_{ijk} x_i y_j z_k=0, h_{ijk} \in \CC \text{.}\end{equation}

The corresponding $X_\varphi$-hypersurface may be smooth or have singularities. Because the variety $X$ is $G$-homogeneous the singular types of the $X_\varphi$-hypersurface and the $X_{g.\varphi}$-hypersurface are the same for any $g\in G$.
In other words the singular type of the $X_\varphi$-hypersurface is a SLOCC invariant of the state $\ket{\varphi}$. Therefore it is natural to look at the entangled classes and characterize them in terms of
the singularities of the corresponding $X_\varphi$-hypersurfaces. In particular if we restrict our attention to those $X_\varphi$-hypersurfaces which have only isolated singularities, then we can use classification results 
from singularity theory.
In particular in the article we show that only a few types of isolated singularities can be obtained by this construction in the case of three-qutrit systems.
We prove:
\begin{theorem}\label{proposition1}
		Let $H_\varphi \cap X$ be a singular hyperplane section of the algebraic variety of separable states for three-qutrit systems, 
		i.e. $X = \mathcal{S}(\PP^2 \times \PP^2 \times \PP^2)\subset \PP^{26}$, defined by a quantum pure state $\ket{\varphi}\in \PP^{26}$. Then $H_\varphi\cap X$ only admits simple or nonisolated singularities. Moreover if  $x$ is an 
		isolated singular point of $H_\varphi \cap X$,  then its singular type is either $A_1$, $A_2$, $A_3$ or $D_4$.
\end{theorem}
The construction employed by the authors was already used to get a finer grained classification of the four-qubit classification by the first author (see \cite{HLP}). This construction is  inspired by 
a construction due to F. Knop \cite{Knop}.
The paper is organized as follows: In Sec. \ref{sing} we recall some basic facts about singularity theory to explain how the calculations leading to Theorem \ref{proposition1} are performed. Then in Sec. \ref{Result} we prove Theorem \ref{proposition1}.
In Sec. \ref{onion} we look at the hierachy of entangled classes given by singular types. Sec. \ref{conclu} is dedicated to concluding remarks.


\section{Simple singularities}\label{sing}
In this section we first recall the definition of simple singularities of complex analytic functions following Arnold's classification of simple singularities \cite{Arnold, SpringerBook1}. Arnold's caracterization of simple singularities
leads to a pratical way of computing the singular type of those as it was done in \cite{HLP}. We then show how to proceed on two examples.
\subsection{Arnold's classification of simple singularities} \label{arnoldClass}

The Segre variety $X=\mathcal{S}(\PP^2\times\PP^2\times\PP^2)\subset \PP^{26}$ being a rationnal algebraic variety, the $X_\varphi$-hypersurfaces will be defined by  complex homogenous polynomials.
The definition (and classification) of simple singularities following Arnold is given in the more general situation of  (germs of) holomorphic map $f:\CC^n\to \CC$.
	\begin{definition}
		A point $x\in \CC^n$ is said to be a \textbf{critical} point of a holomorphic function $f$  if at that point $x$ the derivatives of $f$  equal zero.
	\end{definition}

	\begin{definition}
		A critical point is said to be a \textbf{nondegenerate} or a \textbf{Morse critical }point iff the second differential of the function at that point is a nondegenerate quadratic form i.e. iff the Hessian matrix of $f$ at 
		the critical point is of full rank.
	\end{definition}

	\begin{definition}
		The \textbf{corank} of a critical point of a function is the dimension of the kernel of its second differential at the critical point.
	\end{definition}


	Let us consider now the set $\mathcal{O}_n$ of function-germs at the point 0 $\in \CC^n$. Let $\mathcal{D}_n$ denote the group of germs of biholomorphic maps $g:(\CC^n,0) \to (\CC^n,0)$. This group acts on the space $\mathcal{O}_n$ by the rule $g(f) = f \circ g^{-1}$, where $f\in \mathcal{O}_n$ and $g\in \mathcal{D}_n$. Then, the orbits for this action are exactly the equivalence classes of function-germs. 

	\begin{definition}
		Two function-germs at zero are said to be \textbf{equivalent} if one is taken into the other by a biholomorphic change of coordinates that keeps the point zero fixed.
	\end{definition}

	We can now define a singularity by: 

	\begin{definition}
		Two critical points are said to be \textbf{equivalent} if the function-germs that define them are equivalent. The equivalence class of a function-germ at a critical point is
		called a \textbf{singularity}.
	\end{definition}


\begin{rem}
 From the definition, it follows that the corank of a singular point will be an invariant of the equivalence class defining the singularity. Thus a singularity will  be  nondegenerate or 
 quadratic or of Morse type if and only if its corank is zero. 
\end{rem}

	\begin{definition}
		Two function-germs $f:(\CC^n,0) \to (\CC,0)$ and $h:(\CC^m,0) \to (\CC,0)$ are said to be \textbf{stably equivalent} if they become equivalent after the addition of nondegenerate quadratic forms in supplementary variables: 
$$ f(x_1, \dots, x_n) + x_{n+1}^2 + \dots + x_k^2 \sim h(y_1,\dots,y_m) + y_{m+1}^2 + \dots + y_k ^2$$
	\end{definition}

		This definition allows us to compare degeneracies of critical points of functions of different numbers of variables. Adding quadratic terms of full rank in new variables do not affect the 
		classification of the singular type.

	 ~

	There exists another invariant of singularities usefull to distinguish between singularity types with the same corank. This invariant is the Milnor number, also called the multiplicity 
	of the critical point, and it is defined by introducing the following quotient: 

	\begin{definition}
		The \textbf{local algebra $\mathcal{Q}_f$ of the singularity of f} is defined to be the quotient of the algebra of function-germs by the gradient ideal of $f$: 

$$ \mathcal{Q}_f = \mathcal{O}_n / I_{\nabla f}$$
	\end{definition}

	with $ I_{\nabla f}= \mathcal{O}_n \langle f_1, \dots, f_n \rangle$, the gradient ideal generated by the partial derivatives $f_i = \partial f / \partial x_i$ of the function $f$. 

	\begin{definition}
	The multiplicity of the critical point, i.e. the Milnor number $\mu$ of a singular germ $(f, 0)$,  is equal to the dimension of the local algebra of $(f, 0)$ ,
	$$\mu = dim_{\CC}(\mathcal{Q}_f) =  dim_{\CC}(\mathcal{O}_n / I_{\nabla f} )$$
	\end{definition}

		It can be shown that a critical point is  isolated if, and only if, $\mu < \infty$. 
		
		Simple singularities were defined by Arnold as the class of singularities which are stable in the following sense.
		\begin{definition}
		 A singularity $(f,0)$ is said to be simple if and only if a (sufficiently) small perturbation of the singularity generates only a finite number of non-equivalent classes.
		\end{definition}
		\begin{rem}
		 It is clear from the definition that a  Morse singularity $(f,0)$ is simple as any funtion $f+\varepsilon h$ (for a small $\epsilon$) will be either nonsingular or with Hessian matrix of full rank. 
		\end{rem}

		In his classification of simple singularities \cite{Arnold}, Arnold proved that being simple is equivalent to the following conditions:

	\begin{itemize}
		\item $\mu < \infty$

		\item $corank(\frac{\partial^2 f}{\partial x_i\partial x_j}(x_0)) \leq 2$

		\item if $corank(\frac{\partial^2 f}{\partial x_i\partial x_j}(x_0)) = 2$, then  the cubic term in the degenerate direction of the Hessian matrix is non-zero

		\item if $corank =2$ and the cubic term is a cube, then $\mu < 9$.

	\end{itemize}

	With those conditions Arnold obtained the classification of simple singularities into five different types, two infinite series, $A_n$ and $D_n$, and three 
	exceptional ones, $E_6$, $E_7$ and $E_8$, leading to the well-known classificiation of simple singularities (Table \ref{table}): 

\begin{table}[!h]
		\begin{tabular}{|c|c|c|c|c|}
		  \hline
		$A_n, n\geq 1$ & $D_n, n\geq 4$ & $E_6$ & $E_7$ & $E_8$ \\
		  \hline
		$x^{n+1}$ & $x^2y+y^{n-1}$ & $x ^3 + y ^4$ & $x^3 + xy^3$ & $x^3 + y^5$ \\
		\hline
\end{tabular}
\caption{Normal forms of simple singularities}\label{table}
\end{table}
\begin{rem}
 A function germ $(f,0)$ with a simple singularity is stably equivalent to one of the normal forms of Table \ref{table}.
\end{rem}

\begin{rem}
 According to the notation of Table \ref{table}, Morse singularity are of type $A_1$.
\end{rem}


\subsection{Computing the $X_\varphi$-hypersurface singular type} \label{examples}
We now show on two pratical examples how to calculate the singular type of the $X_\varphi$-hypersurface defined by $\ket{\varphi}$.
	\begin{ex}
		Let $H_{\varphi_1} \in \PP(\mathcal{H^*})$ be a hyperplane, for the linear form $\bra{\varphi_1} = \bra{000} + \bra{111} + \bra{222}$. 
		The hyperplane section $ H_{\varphi_1} \cap X$ is  tangent to $\ket{A} = \ket{021}$. Indeed a tangent vector to $X$ at
		$\ket{A}$ is of the form $\ket{t_1} = \alpha\ket{121} +  \beta\ket{221} +  \delta\ket{001} +  \sigma\ket{011} +  \lambda\ket{020} +  \epsilon\ket{022}$ and we can 
		easly check that $\langle\varphi_1|t_1\rangle = 0$. The function defining $H_{\varphi_1} \cap X$
corresponds to the restriction of the linear form to $X$, which is expressed as: 
	\begin{equation} f_1(x_0,x_1,x_2,y_0,y_1,y_2,z_0,z_1,z_2) = x_0y_0z_0 + x_1y_1z_1 + x_2y_2z_2.\end{equation}

	Besides, $f_1$ can be written in the chart $[0,2,1]$, corresponding to $x_0=y_2=z_1=1$, as a non-homogeneous polynomial in $6$ variables: 

	\begin{equation} f_1(x_1,x_2,y_0,y_1,z_0,z_2) = y_0z_0 + x_1y_1 + x_2z_2.\end{equation}

	Then we  search for singular points of this form, by resolving the two equations $f_1(x_1,x_2,y_0,y_1,z_0,z_2) = 0$ and ${\nabla} f_1(x_1,x_2,y_0,y_1,z_0,z_2) = 0$. This is equivalent  to solve the equations:

	\begin{equation} \left\{\begin{array}{l} 
	y_0z_0 + x_1y_1 + x_2z_2 = 0\\
	y_1= 0, ~ z_2=0, ~ z_0=0\\
	x_1=0, ~ y_0 =0, ~ x_2= 0
	  \end{array}\right.\end{equation}

	We find thus a unique solution that is $x_1=x_2=y_0=y_1=z_0=z_2=0$, which corresponds to $\ket{A} = \ket{021}$ in this chart. The Hessian matrix of $f_1$, at $x=\ket{A}$, is,  

	$$\begin{pmatrix}
   	  0 & 0 & 0 & 1 & 0 & 0  \\
	   0 & 0 & 0 & 0 & 0 & 1  \\  
	 0 & 0 & 0 & 0 & 1 & 0   \\  
	 1 & 0 & 0 & 0 & 0 & 0 \\
  	0 & 0 & 1 & 0 & 0 & 0  \\ 
	0 & 1 & 0 & 0 & 0 & 0 \\
   	\end{pmatrix}$$

which is of rank 6, that is the full rank. So we can conclude that the point $\ket{A} = \ket{021}$ of $H_{\varphi_1}\cap X$ is an isolated singular point of type $A_1$.

	\end{ex}

	\begin{ex}
		Let $H_{\varphi_2} \in \PP(\mathcal{H^*})$ be a hyperplane, for the linear form $\bra{\varphi_2} = \bra{012} + \bra{021} + \bra{102} + \bra{110} + \bra{111} + \bra{200} $. 
		The hyperplane section $ H_{\varphi_2} \cap X$ is  tangent to $\ket{B} = \ket{222}$. Indeed a tangent vector to $X$ at $\ket{B}$ is of the 
		form $\ket{t_2} = \alpha\ket{022} +  \beta\ket{122} +  \delta\ket{202} +  \sigma\ket{212} +  \lambda\ket{220} +  \epsilon\ket{221}$ and we can easly check
		that $\langle\varphi_2|t_2\rangle = 0$. The function defining $H_{\varphi_2} \cap X$
corresponds to the restriction of the linear form to $X$, which is expressed as: 

	\begin{equation}\begin{split} f_2(x_0,x_1,x_2,y_0,y_1,y_2,z_0,z_1,z_2) & = x_0y_1z_2 + x_0y_2z_1 + x_1y_0z_2\\& + x_1y_1z_0 + x_1y_1z_1 + x_2y_0z_0\end{split} \end{equation}

	Besides, $f_2$ can be written in the chart $[2,2,2]$, corresponding to $x_2=y_2=z_2=1$, as a non-homogeneous polynomial in $6$ variables: 

	\begin{equation} \begin{split} f_2(x_0,x_1,y_0,y_1,z_0,z_1) & = x_0y_1 + x_0z_1 + x_1y_0 + x_1y_1z_0\\ &  + x_1y_1z_1 + y_0z_0\end{split}\end{equation}

	Then we search for singular points of this form, by resolving the two equations $f_2(x_0,x_1,y_0,y_1,z_0,z_1) = 0$ and ${\nabla} f_2(x_0,x_1,y_0,y_1,z_0,z_1) = 0$. It is
	equivalent to solve the equations:

	\begin{equation}\left\{\begin{array}{l} 
	x_0y_1 + x_0z_1 + x_1y_0 + x_1y_1z_0 + x_1y_1z_1 + y_0z_0 = 0\\
	   y_1 + z_1 = 0, ~ y_0 + y_1z_0 + y_1z_1 = 0,\\
	   x_1 + z_0 = 0, ~ x_0 + x_1z_0 + x_1z_1 = 0,\\
	   x_1y_1 + y_0 = 0,  ~ x_0 + x_1y_1 = 0
	  \end{array}\right.\end{equation}

	We find  a unique solution that is $x_0=x_1=y_0=y_1=z_0=z_1=0$, which  corresponds to $\ket{B} = \ket{222}$ in this chart. The corresponding Hessian matrix of $f_2$, at $x=\ket{B}$, is 

	$$\begin{pmatrix}
   	  0 & 0 & 0 & 1 & 0 & 1  \\
	   0 & 0 & 1 & 0 & 0 & 0  \\  
	 0 & 1 & 0 & 0 & 1 & 0   \\  
	 1 & 0 & 0 & 0 & 0 & 0 \\
  	0 & 0 & 1 & 0 & 0 & 0  \\ 
	1 & 0 & 0 & 0 & 0 & 0 \\
   	\end{pmatrix}$$

which is of rank 4. One can check, using for instance the software SINGULAR \cite{Singular}, that the Milnor number of this singular hypersurface is $\mu = 4$. So we  conclude that the point $\ket{B} = \ket{222}$ of $H_{\varphi_2}\cap X$ is an isolated singular point of type $D_4$.

	\end{ex}

\section{Proof of Theorem 1}\label{Result}


~

In order to prove Theorem \ref{proposition1}, the computation of the singular type of all $H_\varphi \cap X$ sections is needed. The orbits of $\PP(\mathcal{H})$, under the action of $G$, 
correspond  Nurmiev's classification of $3\times 3\times 3$ matrices under the action of $G$ \cite{Nurmiev}. Moreover, the singularity of an orbit is the same for all the representatives of the latter. According to Nurmiev classification, 
the $G$-orbits of the three-qutrit Hilbert space consist of 5 families (1 family is parameter free --- the nullcone --- and the 4 others depend on parameters). Normal forms for each family are 
 known \cite{Nurmiev}. So, for each Nurmiev's normal form, we compute the corresponding hyperplane section and study its singular type. We focus specifically on singular isolated points. 
We can thus determine the singular type of these points, by using a formal algebraic computing software, putting into practice the steps explained in examples \ref{arnoldClass} and \ref{examples}. The results obtained 
prove Theorem \ref{proposition1}, i.e. only 
$A_1$, $A_2$, $A_3$ and $D_4$ singular types are reached  as shown in Table \ref{nilpotent} and Table \ref{semisimple} where the result of each calculation is provided.

		\begin{table}[!h]
		\begin{tabular}{|c|c|c|}
		  \hline
		\textbf{Orbit} & {\textbf{Hyperplane}} & {\textbf{Singular}} \\
		   &   & \textbf{type} \\
		  \hline \hline
		$N_1$ & $\bra{012}+\bra{021}+\bra{102}$ & $A_3$ \\ 
		 & $+\bra{111}+\bra{120}+\bra{200}$ & \\
\hline
		 $N_2$ & $\bra{012}+\bra{021}+\bra{102}$ & $D_4$ \\ 
		 & $+\bra{110}+\bra{111}+\bra{200}$ & \\
\hline
		$N_3$ & $\bra{002}+\bra{011}+\bra{020}$ & Non- \\ 
		& $+\bra{101}+\bra{112}+\bra{200}$ & isolated\\
\hline
		 $N_4$ & $\bra{002}+\bra{011}$ & Non- \\ 
		& $+\bra{101}+\bra{110}+\bra{220}$ & isolated\\
\hline
		 $N_5$ & $\bra{002}+\bra{020}$ & Non-\\ 
		& $+\bra{021}+\bra{110}+\bra{201} $ & isolated \\
\hline
		 $N_6$ & $\bra{002}+\bra{011}$ & Non-\\ 
		&  $+\bra{101}+\bra{120}+\bra{210} $ & isolated \\
\hline
		 $N_7$
		 & $\bra{002}+\bra{011}$ & Non- \\ 
		&  $ +\bra{020}+\bra{101}+\bra{210}$ & isolated \\
\hline
		 $N_8$
		 & $\bra{002}+\bra{020}$ & Non- \\ 
		&  $ +\bra{111}+\bra{200}$ & isolated \\
\hline
		 $N_9$ & $\bra{000}+\bra{011}$ & Non-\\ 
		&  $ +\bra{111}+\bra{122}$ & isolated \\
\hline
		 $N_{10}$ & $\bra{002}+\bra{011}+\bra{020}$ & Non- \\
		&  $ +\bra{101}+\bra{110}+\bra{200}$ & isolated \\
 \hline
		 $N_{11}$ & $\bra{002}+\bra{020}$ & Non- \\ 
		&  $ +\bra{101}+\bra{210}$ & isolated \\
\hline
		 $N_{12}$ & $\bra{002}+\bra{020}$ & Non-\\ 
		&  $ +\bra{100}+\bra{111}$ & isolated \\
\hline
		 $N_{13}$ & $\bra{002}+\bra{011}$ & Non-\\ 
		&  $ +\bra{020}+\bra{101}+\bra{110}$ & isolated \\
\hline
		 $N_{14}$ & $\bra{002}+\bra{010}$ & Non-\\
		&  $ +\bra{021}+\bra{100}+\bra{201}$ & isolated \\
 \hline
		 $N_{15}$ & $\bra{011}+\bra{022}+\bra{100}$ & Non-\\ 
		&  $ $ & isolated \\
\hline
		 $N_{16}$ & $\bra{002}+\bra{011}$ & Non-\\ 
		& $+\bra{020}+\bra{100} $  & isolated \\
\hline
		 $N_{17}$ & $\bra{001}+\bra{010}$ & Non-\\ 
		& $ +\bra{102}+\bra{120}$  & isolated \\
\hline
		 $N_{18}$ & $\bra{000}+\bra{011}$ & Non-\\ 
		& $+\bra{101}+\bra{112} $  & isolated \\
\hline
		 $N_{19}$ & $\bra{002}+\bra{010}+\bra{101}$ & Non-\\ 
		& $ $  & isolated \\
\hline
		 $N_{20}$ & $\bra{000}+\bra{111}$ & Non-\\ 
		& $ $  & isolated \\
\hline
		 $N_{21}$ & $\bra{001}+\bra{010}+\bra{100}$ & Non-\\ 
		& $ $  & isolated \\
\hline
		 $N_{22}$ & $\bra{000}+\bra{011}+\bra{022}$ & Non-\\ 
		& $ $  & isolated \\
\hline
		 $N_{23}$ & $\bra{000}+\bra{011}$ &  Non-\\ 
		&  & isolated \\
\hline
		 $N_{24}$ & $\bra{000}$ & Non-\\ 
		&  & isolated \\ 
\hline
\end{tabular}
\caption{Hyperplanes and the singular types of the corresponding $X_\varphi$-hypersurfaces for $\ket{\varphi}$ in the nullcone of three-qutrit states. The trivial orbit is omitted.}\label{nilpotent}
\end{table}

For defining more easly each family of Nurmiev's normal forms and the corresponding hyperplanes, let us denote  $X_1 = \bra{000} + \bra{111} + \bra{222}$, $X_2 = \bra{012} + \bra{120} + \bra{201}$ and $X_3 =\bra{021} + \bra{102} + \bra{210}$. Each family is a linear
combination of those linear forms, plus a nilpotent part, and the complex coefficients associated to $X_1$, $X_2$ and $X_3$ must verify a set of conditions, listed as follows: 

	\begin{itemize}

	\item First family: $abc \neq 0$, $(a^3+b^3+c^3)^3 - (3abc)^3 \neq 0$.	

	\item Second family: $b(a^3+b ^3) \neq 0$, $c= 0$.

	\item Third family: $a \neq 0$, $b = c= 0$.

	\item Fourth family: $c = -b \neq 0$, $a= 0$.

	\end{itemize}
\begin{table}[!h]
		\begin{tabular}{|c|c|c|c|}
		  \hline
		 \textbf{Orbits} & {\textbf{Hyperplane}} & {\textbf{Params}} & {\textbf{Sing.}} \\
		   & &  & \textbf{type} \\
		  \hline\hline
		 $F_{1,1}$ & $a.X_1 +  b.X_2 $ & $a$,$b$,$c$ generic &  Smooth \\
		   & $+ c.X_3$ &  &  section\\ 
\hline
		 $F_{2,1}$ & $a.X_1 +  b.X_2$ & $a$,$b$ generic &  $A_1$\\
		   & $+ \bra{021}+\bra{102}$ & $a=0$  & $3A_1$ \\ 
\hline
		 $F_{2,2}$ & $a.X_1 +  b.X_2$ & $a$,$b$ generic &  $2A_1$\\
		   & $+ \bra{021}$ & $a=0$  & $3A_1$ \\ 
\hline
		 $F_{2,3}$ & $a.X_1 +  b.X_2$ & $a$,$b$ generic &  $3A_1$\\
		   & $+\bra{201}$ & $a=0$  & $3A_1$ \\ 
\hline
		 $F_{3,1}$ & $a.X_1+\bra{012}+\bra{021}$ & $a$ generic &  $2A_1$\\
		   & $+\bra{102}+\bra{120}$ &   &  \\ 
\hline
		 $F_{3,2}$ & $a.X_1+\bra{012}$ & $a$ generic &  $3A_1$\\
		   & $+\bra{021}+\bra{102}$ &   &  \\ 
\hline
		 $F_{3,3}$ & $a.X_1+\bra{012}$ & $a$ generic &  $3A_1$\\
		   & $+\bra{021}+\bra{120}$ &   &  \\ 
\hline
		 $F_{3,4}$ & $a.X_1$ & $a$ generic &  $4A_1$\\
		   & $+\bra{012}+\bra{021}$ &   &  \\ 
\hline
		 $F_{3,5}$ & $a.X_1$ & $a$ generic &  $4A_1$\\
		   & $+\bra{012}+\bra{120}$ &   &  \\ 
\hline
		 $F_{3,6}$ & $a.X_1$ & $a$ generic &  $4A_1$\\
		   & $+\bra{021}+\bra{102}$ &   &  \\ 
\hline
		 $F_{3,7}$ & $a.X_1$ & $a$ generic &  $5A_1$\\
		   & $+\bra{012}$ &   &  \\ 
\hline
		 $F_{3,8}$ & $a.X_1$ & $a$ generic &  $5A_1$\\
		   & $+\bra{021}$ &   &  \\ 
\hline
		{}  $F_{3,9}$ & $a.X_1$ & $a$ generic &  $6A_1$\\
\hline
		 $F_{4,1}$ & $b(X_2 - X_3)+\bra{002}$ & $b$ generic &  $A_2$\\
		   & $+\bra{020}+\bra{111}+\bra{200}$ &   &  \\ 
\hline
		 $F_{4,2}$ & $b(X_2 - X_3)+\bra{002}$ & $b$ generic &  $A_3$\\
		   & $+\bra{011}+\bra{020}+\bra{101}$ &   &  \\ 
		   & $+\bra{110}+\bra{200}$ & & \\
\hline
		 $F_{4,3}$ & $b(X_2 - X_3)$ & $b$ generic &  $D_4$\\
		   & $+\bra{000}+\bra{111}$ &   &  \\ 
\hline
		 $F_{4,4}$ & $b(X_2 - X_3)+\bra{001}$ & $b$ generic &  Non-\\
		   & $+\bra{010}+\bra{100}+\bra{200}$ &   & isolated \\ 
\hline
		 $F_{4,5}$ & $b(X_2 - X_3)$ & $b$ generic &  Non-\\
		   & $+\bra{000}$ &   & isolated \\ 
\hline
		 $F_{4,6}$ & $b(X_2 - X_3)$ & $b$ generic &  Non-\\
		   & & & isolated \\
\hline
		\end{tabular}
		\caption{Hyperplanes and the singular types of  the corresponding $X_\varphi$-hypersurfaces where $\ket{\varphi}$ is a state depending on parameters.}\label{semisimple}
		\end{table}

The parameter free family of Nurmiev's classification is called the nullcone and contains only nilpotent orbits. Geometrically the nullcone $\mathcal{N}\subset \PP^{26}$ is  a variety of 
dimension $23$ given by the zero locus of the three generators of the algebra of SLOOC-invariant polynomials on $\CC^3\otimes\CC^3\otimes\CC^3$. In other words the states of the nullcone annilhate all three-qutrit 
invariant polynomials.
See \cite{LT} for a discussion on invariants and covariants of three-qutrit states and also \cite{HLT2} for an analysis of the nullcone in the four-qubit case.

\section{Entanglement classes and onion-like structure of 3-qutrits} \label{onion}
Quantum states SLOCC-equivalent to  $F_{1,1}$ are semisimple states and are the most stable states in the sense that perturbing the state $\ket{F_{1,1}}\rightsquigarrow \ket{F_{1,1}}+\varepsilon\ket{\varphi}$
will not change the family of that state. From our perspective this  can also be understood from the fact that $H_{F_{1,1}}\cap X$ is a smooth hypersurface and that a small perturbation of a 
smooth hypersurface will still give a smooth hypersurface.
In this respect the hypersurface having only simple singularities are the next most stable ones.
Following Arnold's definition of simple singularities, the knowledge of the singular type of a $X_\varphi$-hypersurface provides information about the possible deformations of the singular hypersurface
under a small perturbation. In particular the miniversal deformation of a $D_4$ singularity gives singularities of type $A_3, A_2$ and $A_1$ according to the adjacency diagram (Eq (\ref{adj})).
\begin{equation}\label{adj}
 A_1\leftarrow A_2\leftarrow A_3\leftarrow D_4
\end{equation}
Thus from the entanglement nature of the states corresponding to the normal forms $F_{4,3}$ (resp. $N_2$), we can infer that a small perturbation of the state will lead to states of type $F_{4,2}$ (resp. $N_1$), $F_{4,1}$, $F_{2,1}$ and $F_{1,1}$.

\begin{figure}[!h]
\begin{center}
  \includegraphics[width = 9.0cm]{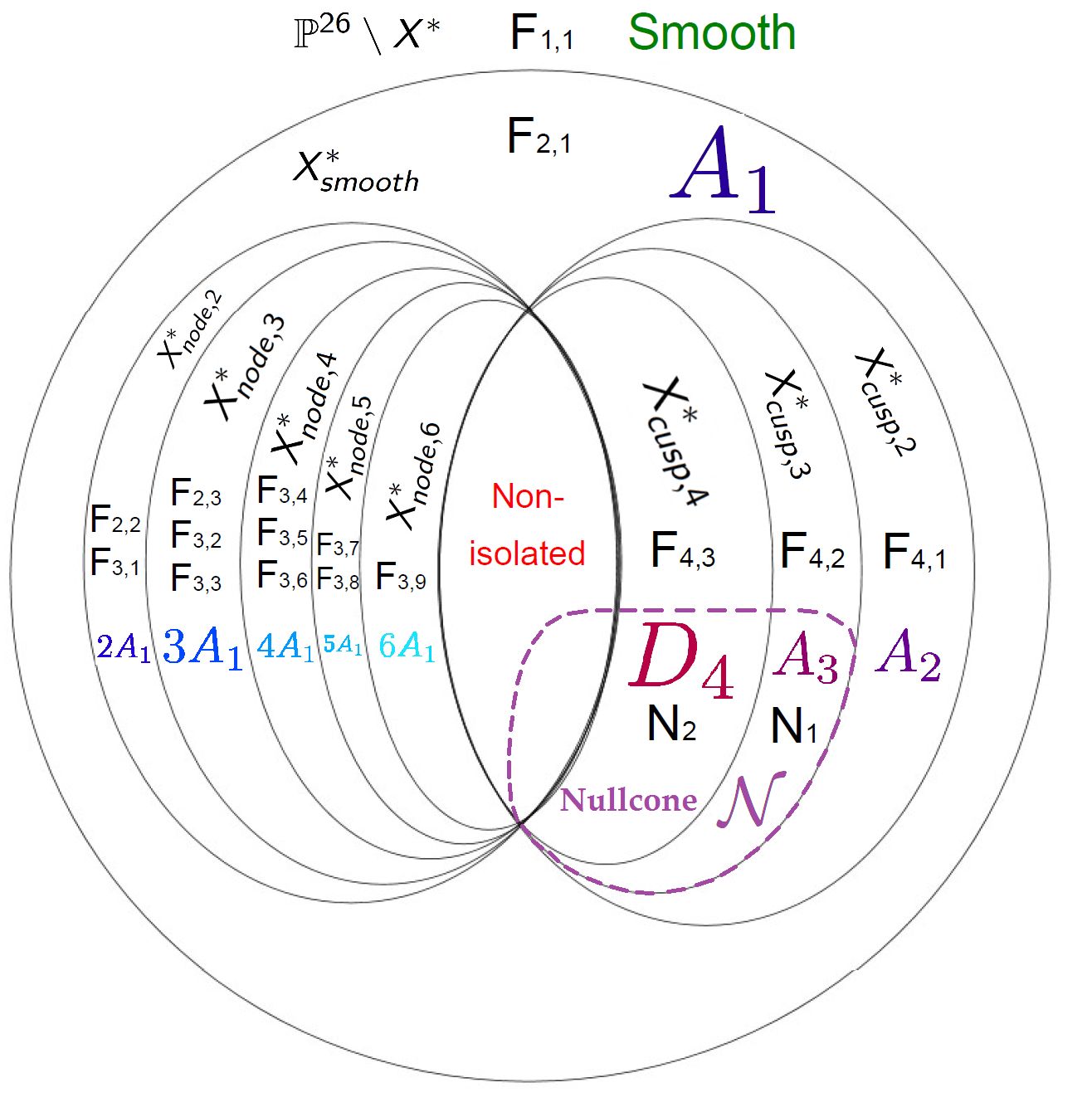}
  \caption{Onion-like structure of the 3-qutrits Hilbert space}\label{qutritentanglement}
\end{center}
\end{figure}

The study of the singular type of the $X_\varphi$-hypersurface can also be reinterpreted in terms of singular locus of the dual variety of $X$. 
Recall that the dual variety $X^*$ is the (algebraic closure of the) set of tangent hyperplanes, i.e.
\begin{equation}
 X^*=\overline{\{H\in \PP(V^*), \exists x\in X, T_x X\subset H\}}
\end{equation}
where $T_xX$ denotes the tangent space of $X$ at $x$. For $X=\mathcal{S}(\PP^2\times\PP^2\times \PP^2)$, the variety $X^*$ is a SLOCC-invariant hypersurface called hyperdeterminant of format 
$3\times 3\times 3$ (see \cite{GKZ} for the theory of hyperdeterminant in general and \cite{Oeding} for the $3\times 3\times 3$ hyperdeterminant in particular).
The hyperdeterminant can be used, as suggested by Miyake \cite{Miyake}, to describe classes of entanglement by looking at the singular locus of this SLOCC-invariant hypersurface.
The smooth points of the hypersurface correspond to hyperplanes defining hyperplane sections with a unique Morse singularity (the hyperplanes which do not belong to the hypersurface
define smooth hyperplane sections, they correspond to states of type $\ket{F_{1,1}}$, as mentioned above, for a general choice of parameters).
Then there are two types of singular points: either they form a node component $X^* _{node}$ (the corresponding hyperplane section has several isolated singular points) or they form a cusp component $X^* _{cusp}$ 
(the singular hyperplane section is not a Morse singularity). Those defintions were introduced by Weyman and Zelevinsky \cite{WZ} and later on used by Miyake in the context of entanglement \cite{Miyake}.
A result of Dimca \cite{dimca} and Parusinski \cite{Paru} insures moreover that the multiplicty of a hyperplane (seen as a point) of the dual variety equals the sum of the Milnor numbers of the corresponding hyperplane section.
Therefore Table \ref{semisimple} tells us that, for instance, the state $\ket{F_{2,3}}$ belongs to $X^* _{node,3}$ (the set of node points of multiplicity $3$) while $\ket{F_{4,2}}$ belongs to $X^* _{cusp,3}$.
Therefore our results allows us to describe the stratification induced by the singular point of $X^*$. It provides an onion-like structure, following the words of Miyake, to describe the three-qutrit entanglement (see Figure \ref{qutritentanglement}).

In this picture the stratification of the ambiant space by SLOCC invariant subvarieties can directely be read from the singular locus of $X^*$.

\section{Conclusion}\label{conclu}
In this paper we employed the classification of simple singularities to look at the three-qutrit entanglement following the approach of \cite{HLP}. This allows us to identify the Nurmiev's normal forms as components 
of the singular locus of the dual variety of the set of separable states. This stratification provides an onion-like description of the classes of entanglement for the three-qutrit case.
Back to the simple singular types which show up in Table \ref{nilpotent} and Table \ref{semisimple}, it is quite surprising to see that the worst isolated singular point is of type $D_4$. It was already the 
case observed in \cite{HLP} for the four-qubit case. However in the four-qubit case the appearence of the $D_4$-type was expected in the sense that the Lie group $SO(8)$, of Dynkin diagram $D_4$, plays a major role in the orbit classification 
of four-qubit entanglement \cite{Verstraete}. Thus in this case the $D_4$ singularity was providing an $ADE$-type correspondence between Lie group and singulariy of the same $ADE$ Dynkin diagram. In \cite{HLP} this correspondence could be clearly
understood from the expression of the $2\times 2\times 2\times 2$ hyperdeterminant, i.e. the dual equation of the variety of separable states, which could be written in the form of the discriminant of the 
miniversal deformation of a $D_4$ singularity. In the case of three-qutrit the correspondence is more mysterious and a direct relation between the expression of the $3\times 3\times 3$ hyperdeterminant, as calculated in \cite{Oeding},
and the discriminant of the minversal deformation of $D_4$ is still missing.

\end{document}